\documentclass[a4paper,fleqn,usenatbib]{mnras}

\usepackage{newtxtext,newtxmath}

\usepackage[T1]{fontenc}
\usepackage{ae,aecompl}

\usepackage{booktabs,caption,fixltx2e}
\usepackage[flushleft]{threeparttable}

\usepackage{graphicx}	
\usepackage{amsmath}	
\usepackage{amssymb}	

\title[The failure of cosmic opacity tests]{The failure of testing for cosmic opacity via the distance-duality relation}

\author[V. Vavry\v{c}uk, P. Kroupa]{
V\'{a}clav Vavry\v{c}uk,$^{1}$\thanks{E-mail: vv@ig.cas.cz}
Pavel Kroupa,$^{2,3}$
\\
$^1$Institute of Geophysics, The Czech Academy of Sciences, Bo\v{c}n\'{i} II, Praha 4, Czech Republic\\
$^2$Helmholtz-Institut f\"ur Strahlen- und Kernphysik, University
        of Bonn, Nussallee 14-16, D-53115 Bonn, Germany\\
$^3$Charles University in Prague, Faculty of Mathematics and 
        Physics, Astronomical Institute, V Hole\v{s}ovi\v{c}k\'ach 2, Praha 8, Czech Republic
}

\date{Accepted 2020 June 30. Received 2020 April 2; in original form 2020 January 18}

\pubyear{2019}

\begin{document}
\label{firstpage}
\pagerange{\pageref{firstpage}--\pageref{lastpage}}
\maketitle
%
\begin{abstract}
%
The distance-duality relation (DDR) between the luminosity distance $D_L$ and the angular diameter distance $D_A$ is viewed as a powerful tool for testing for the opacity of the Universe, being independent of any cosmological model. It was applied by many authors, who mostly confirm its validity and report a negligible opacity of the Universe. Nevertheless, a thorough analysis reveals that applying the DDR in cosmic opacity tests is tricky. Its applicability is strongly limited because of a non-unique interpretation of the $D_L$ data in terms of cosmic opacity and a rather low accuracy and deficient extent of currently available $D_A$ data. Moreover, authors usually assume that cosmic opacity is frequency independent and parametrize it in their tests by a prescribed phenomenological function. In this way, they only prove that cosmic opacity does not follow their assumptions. As a consequence, no convincing evidence of transparency of the universe using the DDR has so far been presented.
\end{abstract}
 
\section{Introduction}

The question whether the Universe is transparent or opaque is of primary importance with fundamental cosmological consequences, because the commonly accepted cosmological model must be revised, if the Universe is partially opaque. Neglecting cosmic opacity might distort the observed evolution of the luminosity density and global stellar mass density in the Universe with redshift \citep{Vavrycuk2017a}. It might partially or fully invalidate the interpretation of the Type Ia supernova dimming as a result of dark energy and the accelerating expansion of the Universe \citep{Aguirre1999b,Aguirre1999a,Aguirre_Haiman2000, Corasaniti2006,Menard2010b, Vavrycuk2019}. The thermal radiation of the intergalactic dust integrated over cosmological history might contribute to the cosmic microwave background (CMB), if intergalactic dust is present, and it can even question some or all of the origin of the CMB as being the relic radiation of the Big Bang \citep{Wright1982, Bond1991, Narlikar2003, Vavrycuk2018}.

The cosmic opacity $\lambda$ is defined as light attenuation $A$ in intergalactic space per unit ray path, caused by light extinction by intergalactic dust grains. It is an integral quantity dependent on the proper dust density, grain-size distribution, and the dust extinction efficiency. Consequently, it is spatially dependent and varies with frequency and redshift \citep{Aguirre1999a, Aguirre1999b, Aguirre2000, Corasaniti2006, Vavrycuk2018, Vavrycuk2019}. Whether the cosmic opacity is appreciable or not can be resolved in several ways. Some authors measured directly the cosmic opacity by dust reddening and found that it is appreciable at close distance from galaxies and in intracluster space \citep{Chelouche2007, Muller2008, Menard2010a}. An averaged V-band attenuation of $\approx$0.03 mag at $z = 0.5$ was measured by \citet{Menard2010a} by correlating the brightness of $\approx$85.000 quasars at $z > 1$ with the position of $24 \times 10^{6}$ galaxies at $z \approx 0.3$ derived from the Sloan Digital Sky Survey. Also, a cosmic opacity $\lambda_V \approx 0.02\, \mathrm{mag \, Gpc}^{-1}$ at $z < 1.5$ is reported by \citet{Xie2015} who analysed the quasar continuum for $\approx$90.000 objects.

The cosmic opacity can also be measured from the hydrogen column densities of Lyman $\alpha$ (Ly$\alpha$) absorbers. In particular, massive clouds with $N_\mathrm{HI} > 2\times 10^{20} \, \mathrm{cm}^{-2}$, called damped Ly$\alpha$ absorbers (DLAs), are self-shielded and rich in cosmic dust being detected in galaxies as well as in intergalactic space \citep{Wolfe2005, Meiksin2009}. Since a relationship between the total hydrogen column density $N_\mathrm{H}$ and the color excess $E(B-V)$ is known:  $N_\mathrm{H} / E(B-V) = 5.6 - 5.8 \times 10^{21} \, \mathrm{cm}^{-2} \, \mathrm{mag}^{-1}$ \citep{Bohlin1978, Rachford2002}, we get $N_\mathrm{H} / A_V = 1.9 \times 10^{21} \, \mathrm{cm}^{-2} \, \mathrm{mag}^{-1}\,$Gpc for  $R_V = A_V / E(B-V) = 3.1$, which is a typical value for our Galaxy \citep{Cardelli1989, Mathis1990}. Taking into account observations of the mean cross-section density of DLAs, $\langle n \sigma \rangle =  \left(1.13 \pm 0.15 \right) \times 10^{-5} \, h \, \mathrm{Mpc}^{-1}$ \citep{Zwaan2005}, the characteristic column density of DLAs, $N_\mathrm{HI} \approx 10^{21} \, \mathrm{cm}^{-2}$, and the mean molecular hydrogen fraction in DLAs of about 0.4-0.6 \citep[their table 8]{Rachford2002}, the cosmic opacity at $z = 0$ is $\lambda_V\approx 1 - 2 \times 10^{-2} \, \mathrm{mag \, Gpc}^{-1}$, which is the result of \citet{Xie2015} based on an analysis of quasar spectra. 

The cosmic opacity is rather small at $z = 0$, but it rapidly increases with redshift and the locally transparent universe might become significantly opaque at high redshifts. The increase of opacity with redshifts is caused through cosmic expansion by an increase of the proper dust density with $z$ as $(1+z)^3$. This is well documented by an observed incidence rate of DLAs in the Ly$\alpha$ forest of spectra of distant quasars \citep{Prochaska_Herbert-Fort2004, Rao2006} and by the effective optical depth calculated from the mean transmitted flux in the Ly$\alpha$ forest \citep{Fan2006, Faucher_Giguere2008, Becker2013}, which steeply grows with redshift. According to \citet[his figure 10a]{Vavrycuk2018}, the visual optical depth could achieve a value of $A_V\approx 0.2\, \mathrm{mag}$ at $z = 1$ and $0.7 - 0.8\, \mathrm{mag}$ at $z = 3$. At $z > 4-5$, the abundance of cosmic dust in the universe is uncertain, but recent papers report dusty galaxies even at $z > 7$ \citep{Watson2015, Laporte2017} and dusty halos around star-forming galaxies at $z = 5-7$ \citep{Fujimoto2019}. Since dust in high-redshift galaxies can efficiently be transported to halos due to radiation pressure as shown by \citet{Hirashita_Inoue2019}, the opacity due to cosmic dust can be appreciable even at redshifts $z > 5-7$. 

Another way how to detect the cosmic opacity is to employ the Etherington's reciprocity relation \citep{Etherington1933}, also known as the distance-duality relation (DDR). The DDR is a general cosmology-independent theorem, which relates the luminosity distance $D_L$ and the angular diameter distance $D_A$. It is based on the assumption of conservation of the number of travelling photons and its violation can indicate the presence of cosmic opacity. This approach was proposed by \citet{Bassett_Kunz_ApJ2004} and so far it has been applied by many authors using various estimations of the distances $D_L$ and $D_A$. The $D_L$ values were mostly estimated using Type Ia supernova (SN Ia) observations \citep{Holanda2010, Lima2011, Li_Lin2018}, and the $D_A$ values were determined, for example, using baryon acoustic oscillations (BAO)  \citep{More2009, Ma_Corasaniti2018}, galaxy clustering in  multipole space \citep{Cooray2006}, the Sunyaev-Zeldovich effect and X-ray brightness of galaxy clusters \citep{Holanda2010, Holanda2013}, ultracompact radio sources \citep{Li_Lin2018} or from galactic-scale strong gravitational lensing systems \citep{Ma2019}. In contrast to direct opacity measurements,  most papers based on the DDR test report negligible cosmic opacity. 

In this paper, we will reexamine the cosmic opacity tests based on the DDR and analyse reasons why most of them indicate negligible opacity of the Universe even though other measurements point to the opposite results. Simulating the quality and extent of currently available datasets, we will replicate the DDR cosmic opacity tests and show how decisive conclusions about the opacity can be achieved. We will discuss other limitations of the DDR approach and propose strategies for getting reliable results.

\section{The distance-duality relation}

This fundamental cosmological theorem relates the luminosity distance $D_L$ with the angular diameter distance $D_A$ as follows:
\begin{equation}\label{eq1}
\frac{D_L}{D_A} \left(1+z\right)^{-2} = 1 \, .
\end{equation}
This equation holds for any cosmological model, in which the redshift $z$ is a measure of the expansion of the Universe and the number of photons travelling along rays in a Riemannian space-time is conserved. Consequently, if DDR is violated, it most likely indicates absorption of photons due to cosmic opacity. Since the flux received by the observer in the opaque universe is reduced by the factor $e^{-\tau (z)/2}$, the observed luminosity distance $D_L^{\mathrm{obs}}$ reads
\begin{equation}\label{eq2}
D_L^{\mathrm{obs}} = D_L e^{\tau (z)/2} \, ,
\end{equation}
where $D_L$ is the luminosity distance for the transparent universe standing in Eq. (1), and $\tau (z)$ is the optical depth of the Universe at redshift $z$. Hence, the DDR is modified for the opaque universe as
\begin{equation}\label{eq3}
\frac{D_L^{\mathrm{obs}}}{D_A} \left(1+z\right)^{-2} = e^{\tau (z)/2} \, ,
\end{equation}
where $\tau (z)$ is usually parametrized in two alternative ways \citep{Li_PhysRevD2013, Liao2013, Ma2019}:
\begin{equation}\label{eq4}
\tau (z) = 2\epsilon z \, ,
\end{equation}
or
\begin{equation}\label{eq5}
\tau (z) = \left(1+z\right)^{2\epsilon} - 1 \, .
\end{equation}
If we are able to determine $D_L^{\mathrm{obs}}$ and $D_A$ in Eq. (3), we can evaluate optical the depth $\tau (z)$ and the parameter $\epsilon$ and perform tests for the cosmic opacity.

\section{Measurements of luminosity distance} 

\begin{figure*}
\includegraphics[angle=0,width=14cm,trim=40 30 40 20, clip=true]{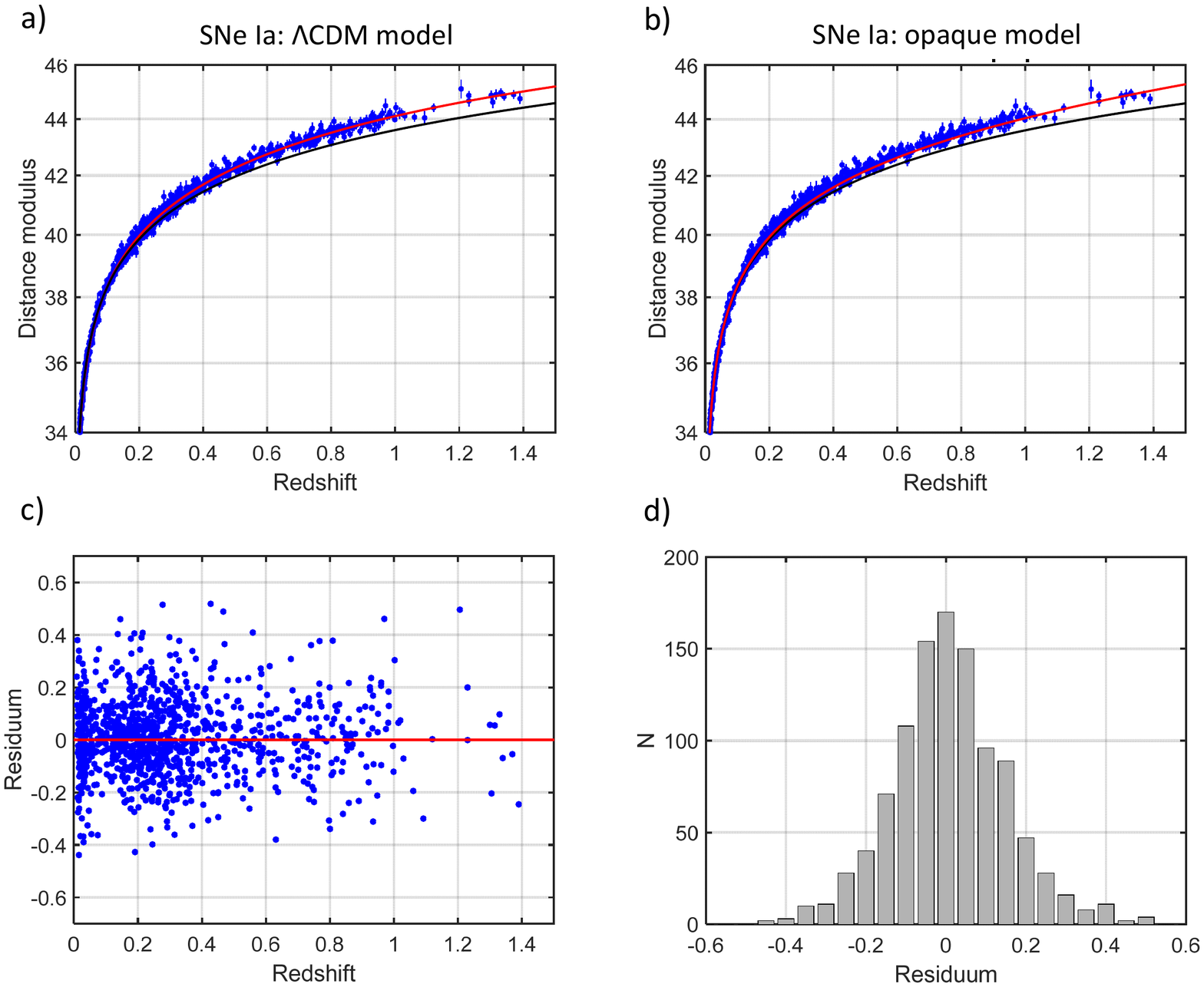}
\caption{
(a-b) Distance modulus as a function of redshift for Type Ia supernova data. (c) Distance modulus residua as a function of redshift and (d) histogram of  distance modulus residua. Blue dots in (a-b) show measurements of the SN Pantheon compilation \citep{Scolnic2018, Jones2018}. The red line in (a-b) shows the $\Lambda$CDM model ($H_0 = 70.9 \, \mathrm{km} \, \mathrm{s}^{-1} \, \mathrm{Mpc}^{-1}$, $\Omega_\Lambda = 0.73$) and the opaque EdS model ($H_0 = 67.2 \, \mathrm{km} \, \mathrm{s}^{-1} \, \mathrm{Mpc}^{-1}$, $\Omega_\Lambda = 0$, $\lambda_B = 0.09 \, \mathrm{mag \, Gpc}^{-1}$), respectively. The black line in (a-b) shows the transparent EdS model with $\lambda_B = 0$. The distance modulus residua in (c-d) are calculated for the $\Lambda$CDM model and appear indistinguishable to those in the opaque EdS model (not plotted here).
}
\label{fig:1}
\end{figure*}

\subsection{Type Ia supernovae observations} 

The most accurate measurement of the luminosity distance $D_L$ as a function of redshift is provided by Type Ia supernova (SN Ia) observations. The luminosity distance of the supernovae was firstly studied by \citet{Riess1998} and \citet{Perlmutter1999}, who revealed that the luminosity of high-redshift SNe Ia declines with redshift more steeply than so far assumed. The observation of the unexpected SN Ia dimming motivated large-scale systematic searches for SNe Ia and resulted in a rapid extension of supernova compilations \citep{Sullivan2011,Suzuki2012,Campbell2013,Betoule2014, Scolnic2018, Jones2018}. The SN Ia dimming was attributed to an accelerating expansion of the Universe, and dark energy was introduced into the cosmological equations. However, several authors pointed out that SN Ia dimming might also be affected or fully produced by light extinction by intergalactic dust \citep{Aguirre1999a, Aguirre1999b, Aguirre_Haiman2000, Inoue_Kamaya2004, Corasaniti2006, Menard2010b, Vavrycuk2019}.

The current supernova compilations Union2.1 \citep{Sullivan2011, Suzuki2012, Campbell2013, Betoule2014, Rest2014, Riess2018} and Pantheon \citep{Scolnic2018, Jones2018} comprise hundreds of SNe Ia discovered and spectroscopically confirmed. The Pantheon dataset is the largest and most accurate SN Ia compilation at present. Every SN Ia is described by its apparent rest-frame B-band magnitude $m_B$, the absolute B-band magnitude $M_B$, the stretch parameter $x_1$, and the colour parameter $c$. These parameters are used for calculating the redshift-dependent distance modulus $\mu(z)$, which serves for testing the cosmological models,
\begin{equation}\label{eq6}
\mu = m_B - M_B + \alpha x_1 - \beta c \,,
\end{equation}
where the coefficients $\alpha$ and $\beta$ are the global nuisance parameters to be determined when seeking an optimum cosmological model. The luminosity distance $D_L$ is related to $\mu$ as follows,
\begin{equation}\label{eq7}
\mu = 25 + 5 \mathrm{log}_{10} D_L^{\rm obs} \,, \,\,
D_L^{\rm obs} = \left(1+z\right)\int^{z}_0\frac{c dz'}{H\left(z'\right)} \,.
\end{equation}
Fig.~\ref{fig:1}a,b shows the SN Ia Pantheon data \citep{Scolnic2018, Jones2018} as a function of redshift together with theoretical curves predicted by two cosmological models: the dark energy model and the opaque universe model (with opacity caused by intergalactic dust). The dark energy model is the standard $\Lambda$CDM model, which is flat and fully transparent and described by the Hubble constant $H_0 = 70.9 \, \mathrm{km} \, \mathrm{s}^{-1} \, \mathrm{Mpc}^{-1}$ and dark energy $\Omega_\Lambda = 0.73$. The opaque universe model is the flat Einstein-de Sitter model (EdS) with a non-zero opacity, being described by the Hubble constant $H_0 = 67.2 \, \mathrm{km} \, \mathrm{s}^{-1} \, \mathrm{Mpc}^{-1}$, dark energy $\Omega_\Lambda = 0$ and the B-band opacity $\lambda_B = 0.09 \, \mathrm{mag \, Gpc}^{-1}$. The model parameters were obtained by fitting the SN Ia Pantheon data using the method of \citet{Vavrycuk2019}. The residua between both theoretical models and the  measurements follow the normal distribution (Fig.~\ref{fig:1}c,d) with a standard deviation of $\sigma = 0.14$. The scatter of measurements is mostly caused by uncertain corrections of the SN Ia dimming due to galactic dust in the Milky Way and in the host galaxy. Since both models fit the data equally well (for details, see \citealt{Vavrycuk2019}), the SN Ia data must be augmented by other independent data for resolving which of the two alternative models is correct. 

\subsection{Gamma-ray bursts and quasars} 
\begin{figure*}
\includegraphics[angle=0,width=14cm,trim=40 30 40 20, clip=true]{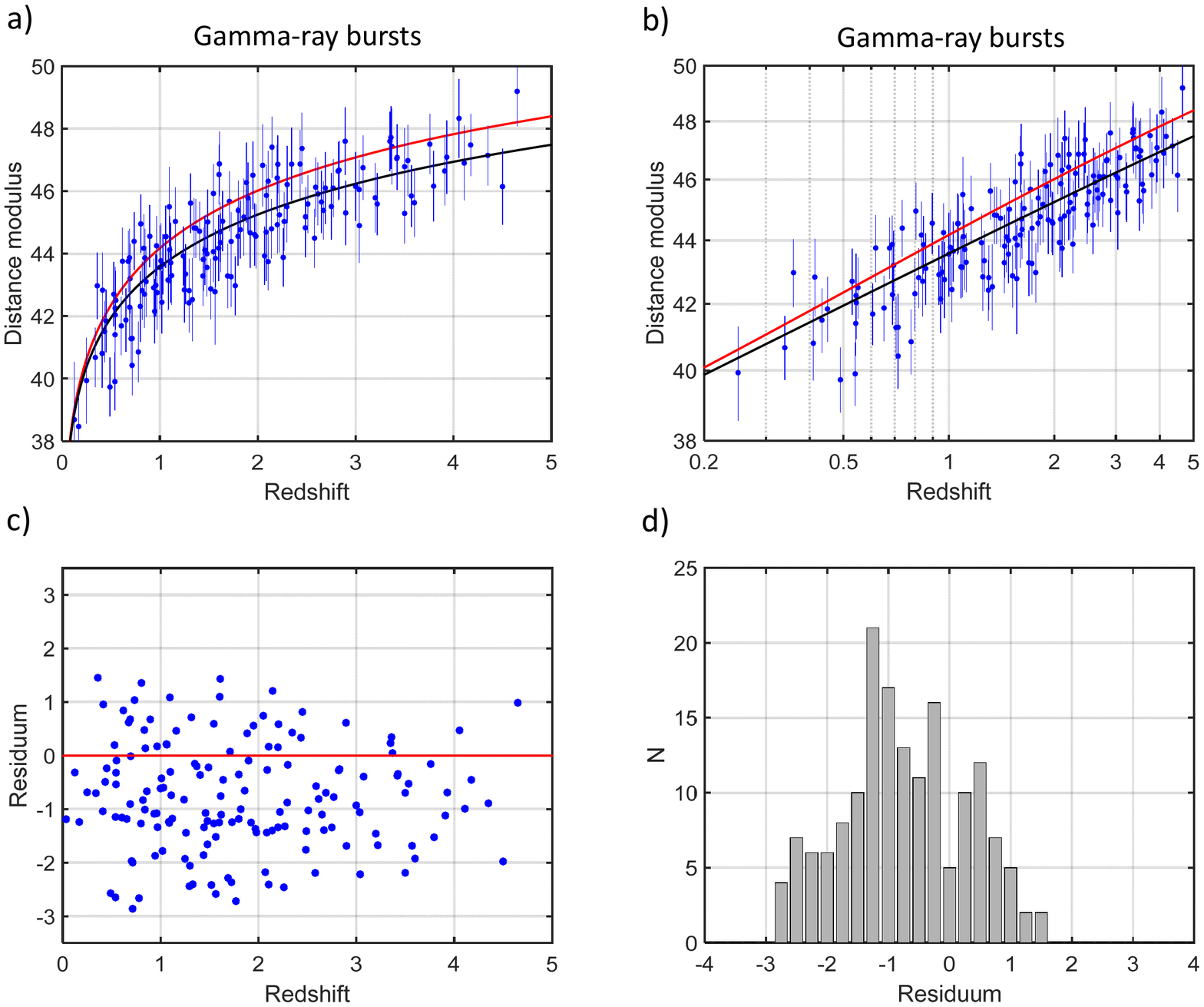}
\caption{
(a-b) Distance modulus as a function of redshift for GRB data in the linear and logarithmic redshift scales. (c) Distance modulus residua as a function of redshift and (d) histogram of  distance modulus residua. Blue dots and error bars in (a-b) show measurements of 139 GRBs \citep{Demianski2017}. The red line in (a-b) shows the $\Lambda$CDM model ($H_0 = 70.9 \, \mathrm{km} \, \mathrm{s}^{-1} \, \mathrm{Mpc}^{-1}$, $\Omega_\Lambda = 0.73$) and the black line in (a-b) shows the transparent EdS model ($H_0 = 67.2 \, \mathrm{km} \, \mathrm{s}^{-1} \, \mathrm{Mpc}^{-1}$, $\Omega_\Lambda = 0$, $\lambda = 0$), respectively. The distance modulus residua in (c-d) are calculated for the $\Lambda$CDM model. The visible bias in (c-d) is removed for the EdS model (not plotted here).
}
\label{fig:2}
\end{figure*}

Since quasars and gamma-ray bursts (GRBs) are the brightest sources in the Universe, they can probe the expansion of the Universe to much higher redshifts (up to $z \approx 7$) than supernovae. Using these sources as standard candles is, however, difficult, because they are extremely variable being characterized by a wide range of luminosity.

GRBs are short and intense pulses of gamma rays produced by a highly relativistic bipolar jet outflow from a compact source \citep{Piran2004}. To calibrate GRBs as standard candles, correlations between various properties of the prompt emission or the afterglow emission were applied \citep{Basilakos_Perivolaropoulos2008}. For example, the correlation between the peak photon energy $E_p$ and the isotropic equivalent radiated energy $E_{iso}$ is employed and the $E_p - E_{iso}$ correlation is calibrated using some other independent data. 

Fig.~\ref{fig:2}a,b shows a dataset of 139 GRBs reported by \citet{Demianski2017} as a function of redshift ranging from 0 to 5. The GRBs are calibrated by SN Ia data and shown together with theoretical curves describing the $\Lambda$CDM model and the transparent EdS model. The residua between data and the $\Lambda$CDM model follow the normal distribution (Fig.~\ref{fig:2}c,d) with the mean significantly departing from zero. The standard deviation is $\sigma = 1.11$, which is about 8 times higher than that for the SN Ia data. Even though the scatter of GRBs is quite high, the EdS model fits the data visibly better than the $\Lambda$CDM model. 

However, if the same dataset is calibrated in a different way, the redshift dependence of $D_L$ might be different. For example, \citet{Amati2019} used observational Hubble data for the calibration \citep{Moresco2016} and the difference between the predictions of the $\Lambda$CDM model and the GRB observations was reduced. This points to an essential weakness of the GRBs data - their large uncertainties produced by calibration, which introduces a so-called 'circularity problem' \citep{Kodama2008, Amati2019}. Since a sufficient number of low-redshift GRBs is missing, the correlation between radiated energy (or luminosity) and the spectral properties is established under a presumed cosmology. Obviously, the benefit of the GRB data is low after such a calibration, because GRBs just replicate the behaviour of the calibration data.

As regards quasars, several approaches considering quasars as standard candles have been proposed but they also suffer from a high scatter of the luminosity-related relation \citep{Bisogni2017}. A suggested method for estimating quasar distances is based on a known non-linear relation between luminosities in the X-rays and UV bands \citep{Lusso_Risaliti2016, Lusso_Risaliti2017}. However, the analysis of the quasar dataset reported by \citet{Risaliti_Lusso2019}, which consists of $\approx$1600 selected sources obtained by cross-correlating the SDSS-DR7 and SDSS-DR12 quasar samples with 3XMM-DR7 X-ray detections, reveals a scatter even larger than that of the GRBs \citep[their fig. 2]{Lusso2019}. Consequently, the currently available quasar data are almost useless for testing for opacity using the DDR. 

\section{Measurements of angular diameter distance}

\subsection{Galaxy clusters}

\begin{figure*}
\includegraphics[angle=0,width=14cm,trim=40 30 40 20, clip=true]{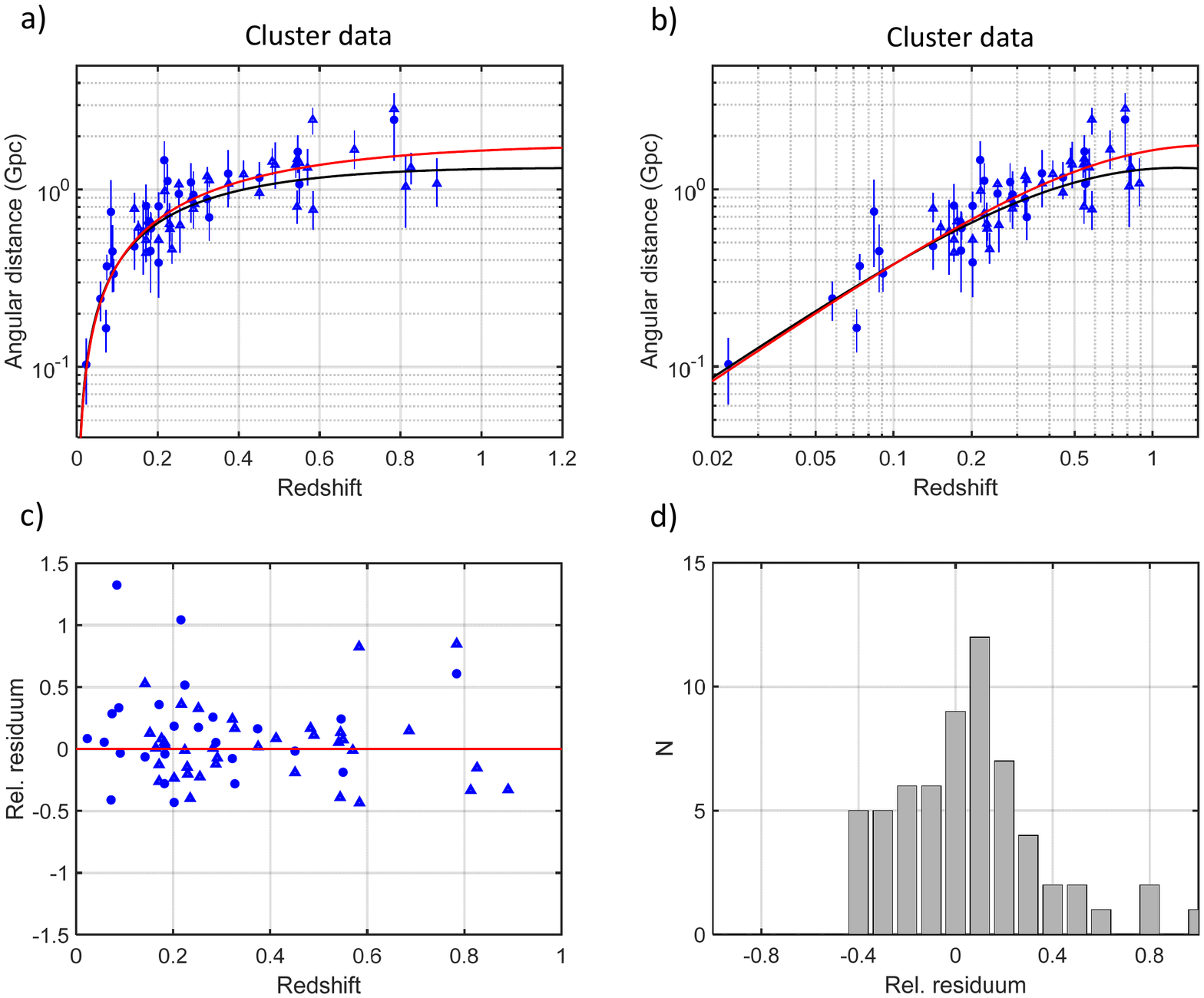}
\caption{
(a-b) Angular diameter distance $D_A$ as a function of redshift for 25 galaxy clusters (blue dots) published by \citet{DeFilippis2005} and for 38 galaxy clusters (blue triangles) published by \citet{Bonamente2006}. (a) Lin-log scale, (b) log-log scale. (c) Relative residua of $D_A$ displayed in (a-b) as a function of redshift, and (d) histogram of the relative residua. The red and black lines in (a-b) show the $\Lambda$CDM model and the opaque EdS model, respectively. The residua in (c-d) are calculated for the $\Lambda$CDM model. The residua for the EdS model are very similar to those for the $\Lambda$CDM model (not plotted here).
}
\label{fig:3}
\end{figure*}

The measurements of the angular diameter distance $D_A$ are more involved and less accurate than those of the luminosity distance $D_L^{\rm obs}$. For example, \citet{DeFilippis2005} and \citet{Bonamente2006} determine $D_A$ of galaxy clusters using data from the Sunyaev-Zeldovich (SZ) effect and the X-ray data. The method is based on observations of the X-ray surface brightness and measurements of the SZ temperature decrement due to the scattering of electrons in galaxy clusters on the CMB \citep{Birkinshaw1999}. The results depend on the mass model and shape of galaxy clusters, which introduce significant uncertainties \citep{Meng2012}. 

In order to demonstrate how measurements of $D_A$ fit cosmological models, we use galaxy cluster data reported by \citet{DeFilippis2005} and \citet{Bonamente2006}, which  cover redshifts up to 1 (Fig.~\ref{fig:3}a,b). The relative residua of measurements from the model are rather high and achieve values up to $\pm$40\%, the standard deviation being $\pm$32\% (Fig.~\ref{fig:3}c,d). The differences in $D_A$ for the $\Lambda$CDM model and for the opaque EdS model are quite small for redshifts up to $z \approx 0.5$. Both models deviate visibly at $z > 0.5$, but the scatter of measurements in this redshift interval is so high that the data cannot resolve a preferable  model (Fig.~\ref{fig:3}a,b). Consequently, it is obvious that when using the SN Ia data for calculating $D_L^{\rm obs}$ (Fig.~\ref{fig:1}) and galaxy cluster data for calculating $D_A$ (Fig.~\ref{fig:3}), no decisive conclusions about cosmic opacity can be made regardless of the published results of \citet{Li_ApJL2011}, \citet{Li2013}, \citet{Yang2013} and others.

\subsection{Large galaxy surveys and BAO}

Another possibility is to measure $D_A$ from angular clustering spectra of galaxies and baryon acoustic oscillations (BAO) \citep{Cooray2006, Beutler2011, Kazin2014}. The method is based on associating the clustering spectrum with known physical scales such as the sound horizon of the last scattering surface calibrated through the CMB anisotropy. Such an approach needs, however, data from large galaxy surveys and provides results at a very limited number of low redshifts only ($z<1$). For example, \citet{Ma_Corasaniti2018} calculate $D_A$ at effective redshifts 0.44, 0.60 and 0.73 using the WiggleZ survey \citep{Blake2012} and at effective redshifts 0.38, 0.51 and 0.61 using the Baryon Oscillation Spectroscopic Survey (BOSS) DR12 \citep{Alam2017}. Also, this approach depends on the cosmic matter density and the Hubble constant, hence it cannot be considered as cosmological-model independent. Moreover, the BAO methods ignore the impact of cosmic dust on the properties of the CMB \citep{Vavrycuk2017b}.

\subsection{Strong gravitational lensing}

\citet{Liao2016} proposed to determine $D_A$ using strong gravitational lensing systems, see also \citet{Holanda2017, Fu_Li2017}. Based on an assumption about the lens mass density profile and on knowledge of the Einstein radius, stellar central velocity dispersion, and redshifts of lens and the source, $z_l$ and $z_s$, respectively, the method provides the angular diameter distance ratio $R_A (z_l,z_s) = D_A^{ls}/D_A^s$, where $D_A^{ls}$ and $D_A^s$ are the angular diameter distances from lens to the source and from the observer to the source, respectively. The inaccuracies in $R_A$ are produced by an estimate of the Einstein radius, which depends on the mass profile of lenses, and by the observer-lens distance, which must be calculated under some cosmological model. Moreover, the DDR must be modified in order to compare distance ratios $R_A$ and $R_L = D_L^l/D_L^s$ instead of comparing $D_A$ and $D_L$. As a consequence, the resolution of the opacity test is low, because calculating the distance ratio $R_L$ is numerically unstable and introduces large errors. 

Some of the above-mentioned deficiencies are avoided by additional measurements of the time-delay distance, which is a combination of three angular diameter distances, $D_A^{\Delta t} = (1+z^l) D_A^l D_A^s/D_A^{ls}$, where $l$ and $s$ stands for lens and source, respectively \citep{Treu_Marshall2016, Suyu2017, Jee2016}. If we multiply the angular diameter distance ratio $R_A$ by the time-delay distance $D_A^{\Delta t}$, we get the angular diameter distance of lens $D_A^l$ being completely independent of the source redshift. As regards uncertainties of $D_A^l$, \citet{Yildirim2020} report $\approx 3$\% precision for best cases, but it is probably more realistic to use an average of 5-10\% precision for reasonably extensive datasets in future studies \citep{Jee2016, Liao2019}. However, the currently reported datasets display much higher uncertainties \citep{Rana2017}.

\subsection{Ultracompact radio sources}

\begin{figure*}
\includegraphics[angle=0,width=14cm,trim=40 30 40 20, clip=true]{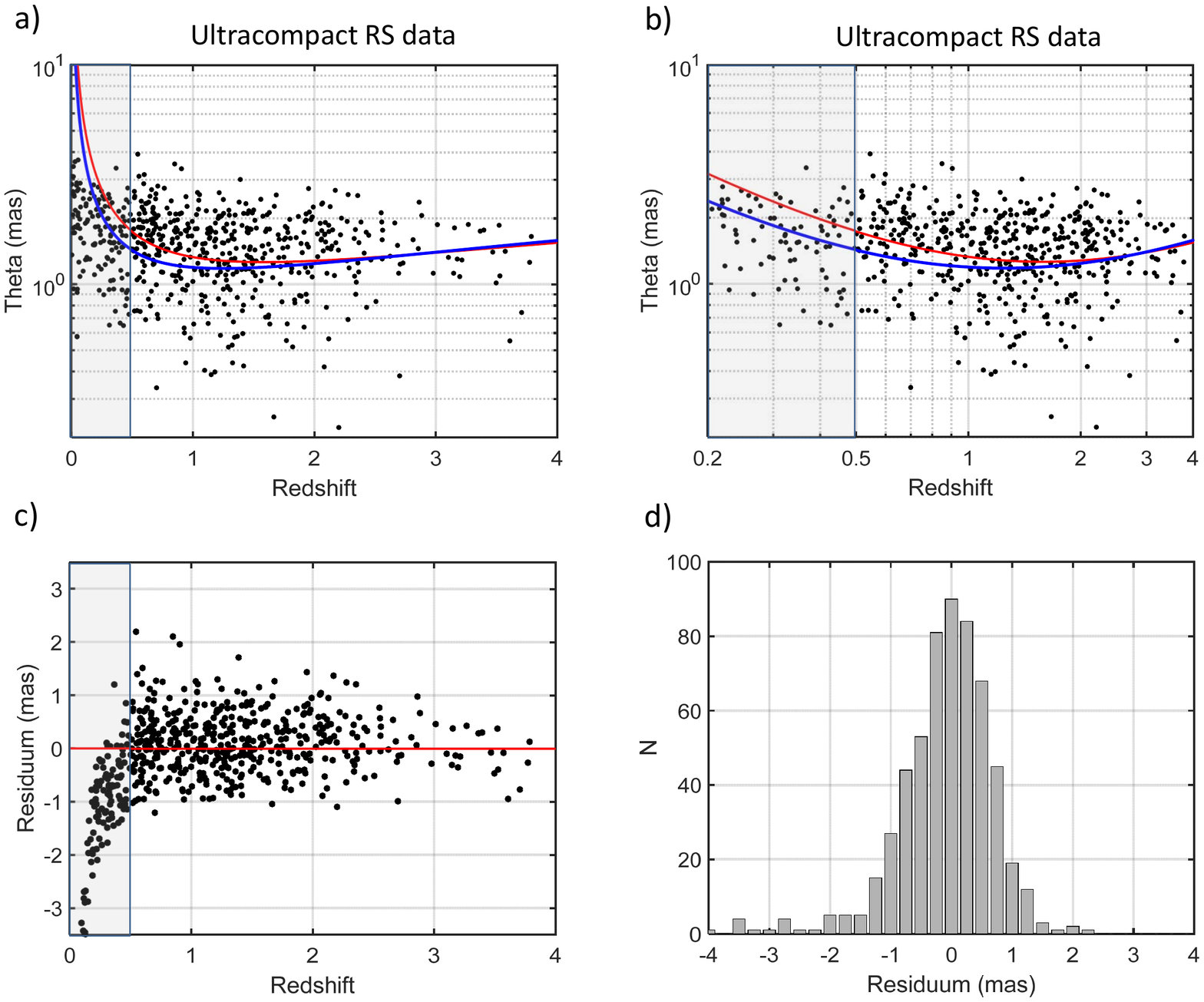}
\caption{
Angular size $\theta$ of ultracompact RSs as a function of redshift. (a-b) Dataset of 613 ultracompact RSs (black points) taken from \citet[see \url{http://nrl.northumbria.ac.uk/13109/}]{Jackson2012} is shown together with functions for the $\Lambda$CDM model (red line, $d_0 = 11$ pc) and the (transparent or opaque) EdS model (blue line, $d_0 = 7.5$ pc) in the linear and logarithmic redshift scales. (c) Relative residua of $\theta$, and (d) histogram of the relative residua. The residua in (c-d) are calculated for the $\Lambda$CDM model. The shaded area marks measurements affected by the selection bias \citep{Jackson2012}.
}
\label{fig:4}
\end{figure*}

The angular diameter distance $D_A$ can also be measured using ultracompact radio sources (RSs) with angular sizes of order of milliarcseconds, which could be measured by a very-long-baseline interferometry (VLBI) at $z < 4-5$. Since \citet{Kellermann1993} showed on 79 ultracompact RSs associated with active galaxies and quasars that their linear size $d_0$ is approximately redshift independent, the ultracompact RSs became potential candidates for the cosmological rulers (\citealt{Gurvits1994,Jackson_Jannetta2006, Jackson2012}). The angular diameter distance is obtained from the observed angular size $\theta$ as $D_A = d_0/\theta$, where the linear size $d_0$ is not known \citep{Gurvits1999, Li_Lin2018}. As the RSs are usually located at different redshifts than SNe Ia used for calculating $D_L$, the distance modulus of RSs must be recalculated to match the redshift of each SN. 

However, even this method is not optimal; it is assumed that measurements at $z < 0.5$ suffer from a strong selection bias \citep[their fig. 1]{Li_Lin2018}, and those at higher redshifts are highly scattered (see Fig.~\ref{fig:4}) with an uncertainty of $\pm 60 \%$. Moreover, they are roughly redshift independent and might easily be distorted by  a systematic error introduced by the unknown linear size $d_0$. For example, \citet{Cao2017} determined $d_0$ using Hubble parameter measurements based on cosmic chronometers \citep{Jimenez_Loeb2002} and obtained $d_0 = 11.0 \pm 0.3$ pc.
However, similarly as for the GRBs, calibrating ultracompact RSs by independent data is dangerous and not self-consistent, because we introduce the circularity problem and a bias in favour of calibration data. 
 
The resolution of raw ultracompact RSs is demonstrated in Fig.~\ref{fig:4}a,b, which shows the $\theta - z$ dependence for 613 ultracompact RSs reported by \citet[see \url{http://nrl.northumbria.ac.uk/13109/}]{Jackson2012} together with functions predicted by the $\Lambda$CDM model (red line, $d_0 = 11$ pc) and the EdS model (blue line, $d_0 = 7.5$ pc). Both the functions are almost indistinguishable for $z > 1$. They start to deviate for $z < 1$, where the EdS model seems to fit the data slightly better. However, the data for $z < 0.5$ are considered as unreliable because of a selection bias. Hence, ultracompact RSs themselves do not favour the $\Lambda$CDM model and the observational radio data cannot be used to effectively derive cosmological information as incorrectly claimed by \citet{Cao2017} or by \citet{Cao2018}. Also, we can speculate that the threshold of $z = 0.5$, dividing biased and unbiased RSs data due to the selection effect, was chosen just for removing discrepancies between the observations and the $\Lambda$CDM model.

\section{Numerical modelling}

\begin{figure*}
\includegraphics[angle=0,width=14cm,trim=40 70 40 80, clip=true]{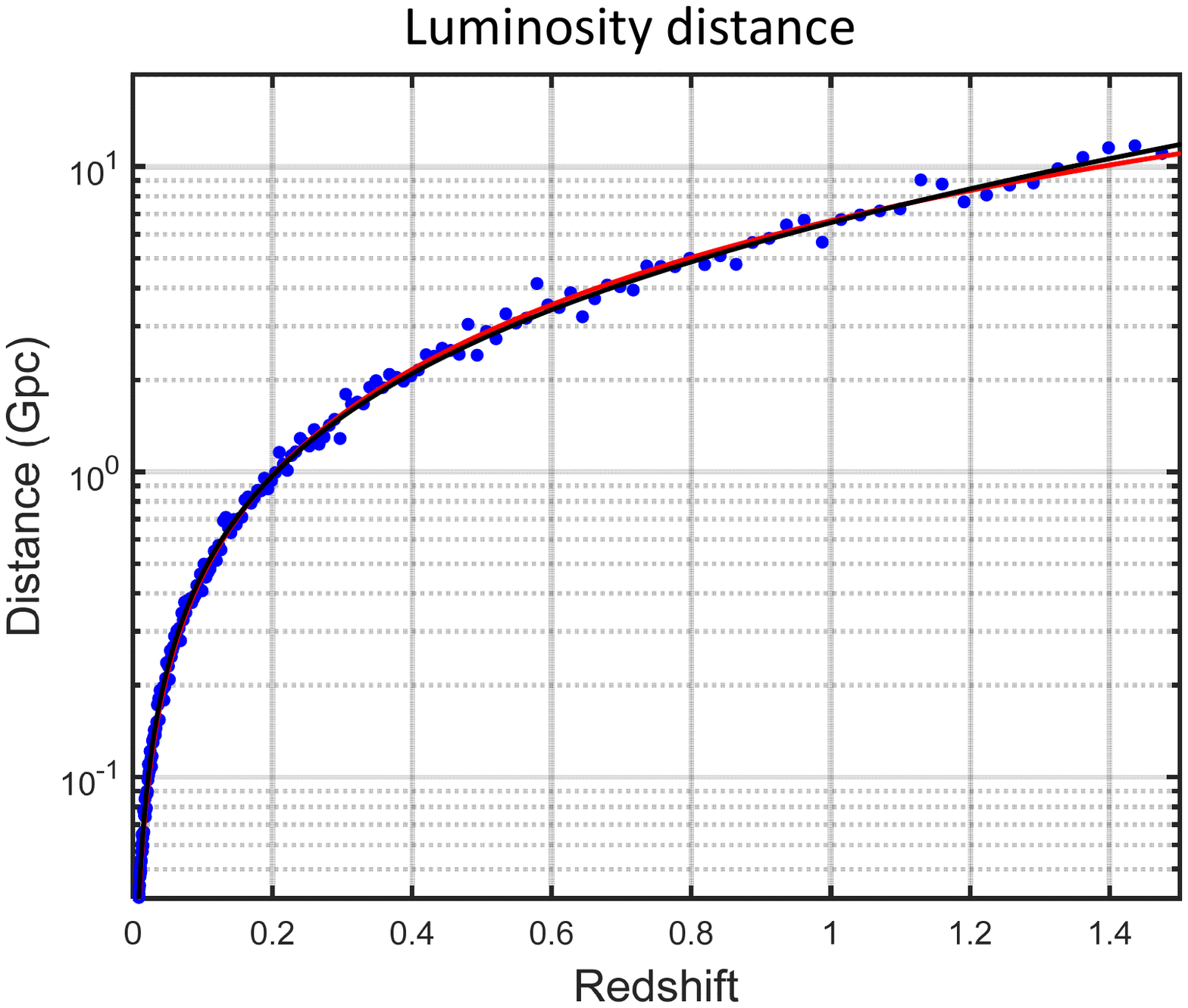}
\caption{
Synthetic $D_L^{\rm obs}$ data simulating the SN Ia observations. The red line - the $\Lambda$CDM model, the black line - the opaque EdS model. The residuals of synthetic data are comparable with those for the SN Ia observations (Fig.~\ref{fig:1}d).
}
\label{fig:5}
\end{figure*}

\begin{figure*}
\includegraphics[angle=0,width=14cm,trim=40 30 40 20, clip=true]{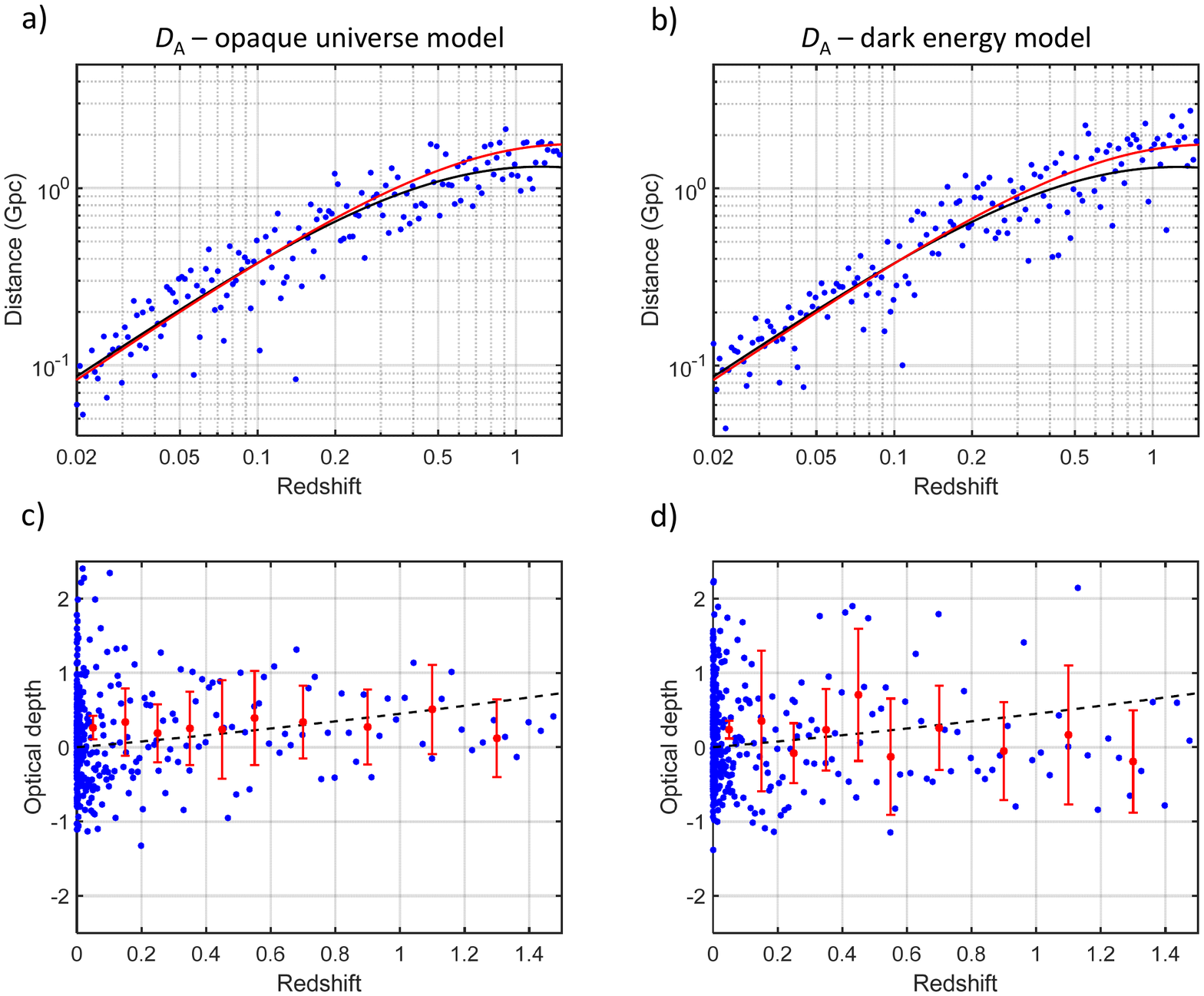}
\caption{
(a-b) Synthetic $D_A$ data simulating galaxy cluster observations, and (c-d) the calculated optical depth using the DDR. The $D_A$ data are generated for the opaque EdS model in (a) and for the $\Lambda$CDM model in (b). (c) Optical depth calculated from the $D_A$ data in (a). (d) Optical depth calculated from the $D_A$ data in (b). The black line - the opaque EdS model, the red line - the $\Lambda$CDM model. Blue points in (c-d) - the individual values of calculated optical depth; red points with error bars in (c-d) - the binned optical depth with the 95\% confidence intervals corresponding to the following redshift bins: [0, 0.1, 0.2, 0.3, 0.4, 0.5, 0.6, 0.8, 1.0, 1.2, 1.4]. The black dashed line - the true optical depth. 
}
\label{fig:6}
\end{figure*}

\begin{figure*}
\includegraphics[angle=0,width=14cm,trim=40 30 40 20, clip=true]{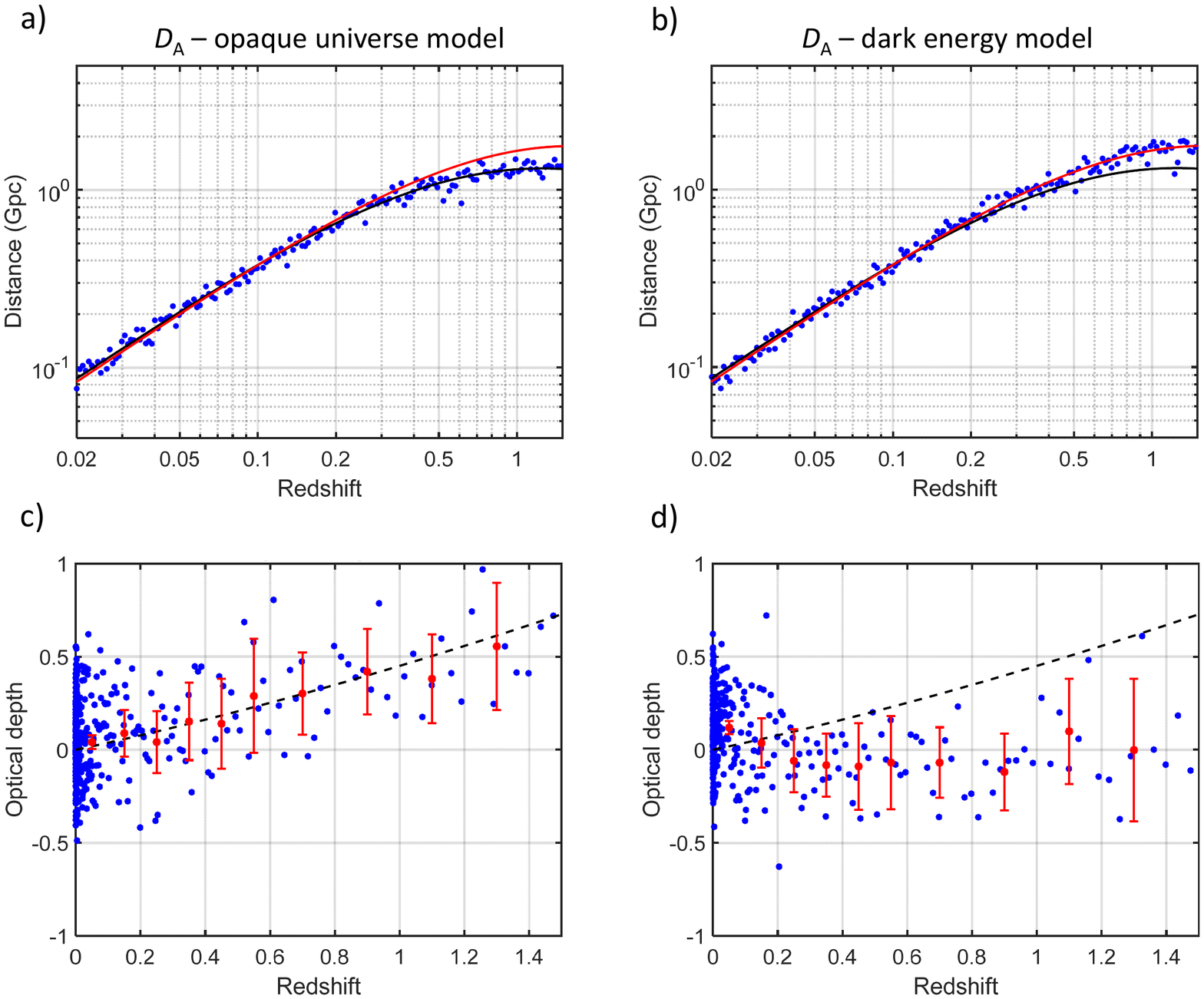}
\caption{
Synthetic $D_A$ data simulating anticipated strong lensing observations. The $D_A$ data display a four times lower noise level than those in Fig.~\ref{fig:6}. For other details, see the caption of Fig.~\ref{fig:6}.
}
\label{fig:7}
\end{figure*}

Since Figs~\ref{fig:1}-~\ref{fig:4} demonstrate essential uncertainties in the data and their redshift coverage, we perform modelling with synthetic data of various quality and probe the power of the DDR for detecting cosmic opacity. The goal of the tests is not just to check the resolution power of existing data, but also to define requirements on data quality for successfully resolving the opacity problem in the future. For the purpose of this test, we assume that the 'true' model is the opaque EdS model described by the Hubble constant $H_0 = 67.2 \, \mathrm{km} \, \mathrm{s}^{-1} \, \mathrm{Mpc}^{-1}$ and $\Omega_\Lambda = 0$. Since the cosmic opacity is strongly frequency dependent, its value is chosen according to the specific type of simulated $D_L$ data. We calculate 'true' luminosity and angular diameter distances, $D_L^{\rm obs}$ and $D_A$, respectively, for the same objects and contaminate them by noise to simulate uncertainties in measurements. We perform tests using data characterized by two different redshift intervals: low-redshift data with $z < 1.5$, and high redshift data with $z < 5$.

\subsection{Low-redshift data (\boldmath{$z < 1.5$})}

The low-redshift synthetic $D_L$ data mimic the current SN Ia Pantheon data, because this dataset provides $D_L$ with the highest accuracy in this redshift range. The level of errors is fixed, because we do not expect the quality of the SN IA data to be essentially improved in the near future. The B-band opacity of the fiducial model is $\lambda_B = 0.09 \, \mathrm{mag \, Gpc}^{-1}$ and fits the SN Ia Pantheon dataset. The fitting procedure is described in \citet{Vavrycuk2019}, where the SN Ia Union2.1 dataset was analyzed. 

The errors in $D_A$ have two alternative levels in order to simulate uncertainties in the currently available galaxy cluster data (uncertainty level of 32\%) and those in strong lensing data available possibly in the near future (uncertainty level of 8\%). The datasets are rather dense with 360 points, which cover the redshift interval $0 < z < 1.5$ typical for the SN Ia measurements. The density of points decreases with redshift. Covering the whole redshift interval by observations allows us to calculate the optical depth $\tau(z)$ from $D_L$ and $D_A$ using Eq. (3) at individual redshifts and to average $\tau(z)$ within redshift bins. 

Since some methods of measuring $D_A$ might not be fully independent of the cosmological model due to the circularity problem (e.g. the galaxy clustering method, the BAO method, gravitational lensing, ultracompact RSs), we also consider inversions for optical depth $\tau(z)$ using $D_L$ corresponding to the 'true' cosmological model (opaque EdS model) but $D_A$ corresponding to another cosmological model ($\Lambda$CDM model) introduced by a biased calibration procedure.

Figs~\ref{fig:5} and ~\ref{fig:6}a,b show $D_L$ from simulated SN Ia data and $D_A$ from simulated galaxy cluster data, and Fig.~\ref{fig:6}c,d shows the resultant optical depth $\tau(z)$. The true optical depth increases with redshift (black dashed line in Fig.~\ref{fig:6}c,d), but no such trend is observed for individual points of $\tau(z)$ as well as for its binned values. Hence, simulated SN Ia data and galaxy cluster data fail to detect the opacity, even though the redshift coverage was much better for simulated galaxy clusters than for real observations (compare Fig.~\ref{fig:2} and Fig.~\ref{fig:6}). The real observations of galaxy clusters display gaps in redshift and are limited to redshifts $z < 1$, for which the optical depth is rather low and less detectable.

Next, the accuracy of the $D_A$ measurements is increased 4 times to simulate uncertainties anticipated in future strong lensing data \citep{Liao2019}. If both $D_L$ and $D_A$ are determined under no a priori assumption of the cosmological model (Figs~\ref{fig:5} and ~\ref{fig:7}a), the data reveal the true increasing tendency in the retrieved optical depth $\tau(z)$ (Fig.~\ref{fig:7}c). However, the non-zero optical depth is statistically significant only for redshifts $z \gtrapprox 0.6$. Moreover, if $D_A$ is biased because of an implicit assumption of the $\Lambda$CDM model (Fig.~\ref{fig:7}b), the retrieved optical depth $\tau(z)$ is incorrectly close to zero with no increasing tendency characterizing the true $\tau(z)$. 

We emphasize that the extent and accuracy of synthetic $D_A$ data shown in Fig.~\ref{fig:7} are quite high: the $D_L$ and $D_A$ correspond to the same objects and cover densely the redshift interval $0 < z < 1.5$. This data quality is much higher than that of currently available observations, but it might mimic the quality of future strong lensing data \citep{Liao2019}. Nevertheless, even in this case, the DDR method yields very approximate values of $\tau(z)$ with a rather high scatter in individual redshift bins. Moreover, if $D_A$ measurements are biased because of assuming some cosmological model when calibrating $D_A$, the DDR might fail and incorrectly yield a zero cosmic opacity even for the true opaque universe (Fig.~\ref{fig:7}d).

\subsection{High-redshift data (\boldmath{$z < 5$})}

\begin{figure*}
\includegraphics[angle=0,width=16cm,trim=40 80 40 40, clip=true]{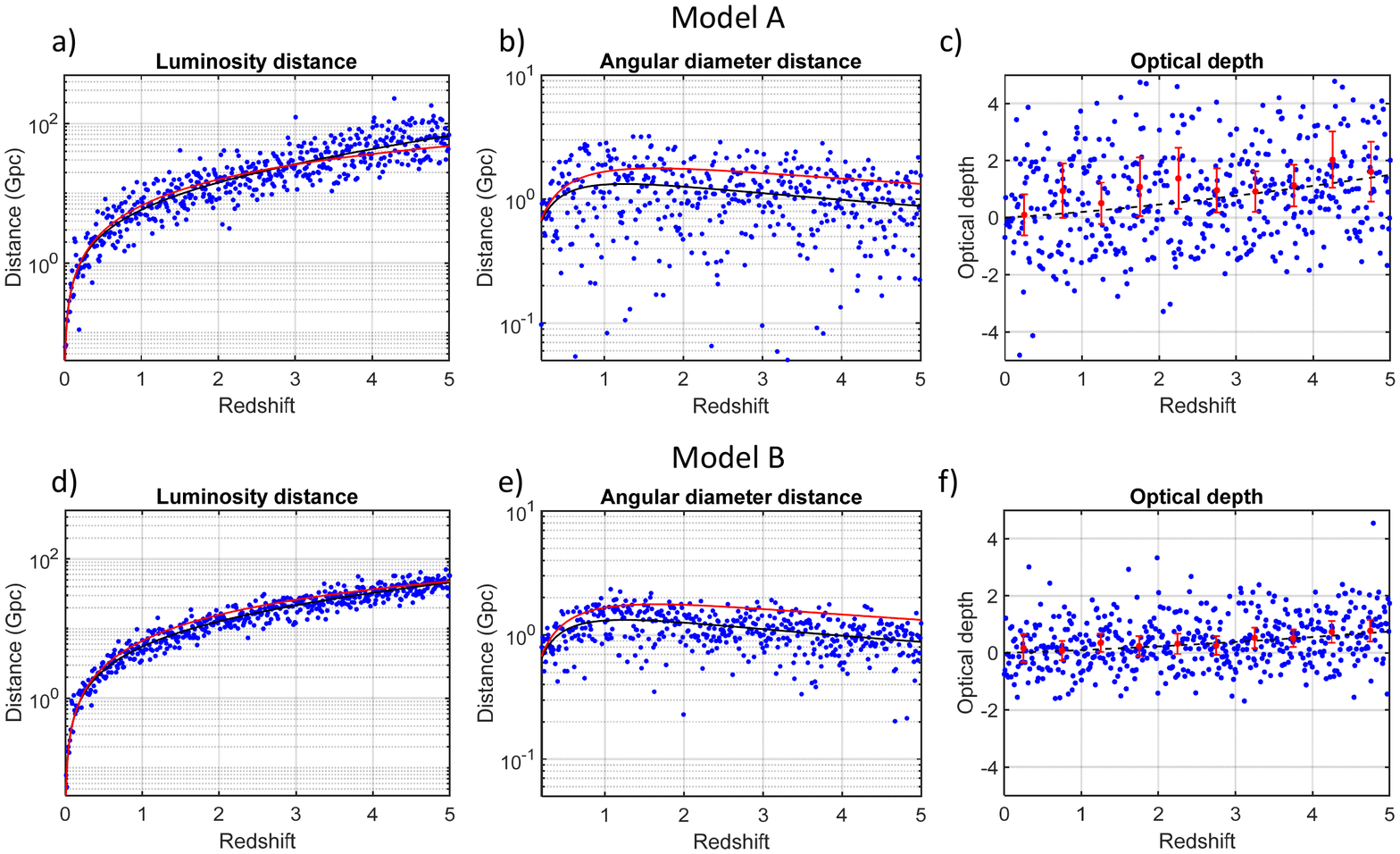}
\caption{
Synthetic high-redshift $D_L$ and $D_A$ data and optical depth in model A (a-c) and model B (d-f). The $D_L$ and $D_A$ data are calculated for the opaque EdS model. The black line in the $D_L$ and $D_A$ plots - the opaque EdS model, the red line in the $D_L$ and $D_A$ plots - the $\Lambda$CDM model. (c,f) Blue points - individual values of calculated optical depth; red points with error bars - the binned optical depth with the 95\% confidence intervals corresponding to bins in the redshift interval of (0, 5) with step of 0.5. The black dashed line - the true optical depth. 
}
\label{fig:8}
\end{figure*}

If we use quasars or GRB data for determining $D_L$ and ultracompact RSs for determining $D_A$, the DDR test can be applied to redshifts up to 5. Since the cosmic opacity should increase with redshift, its detection at high redshifts should be easier and uncertainties in observed $D_L$ and $D_A$ could be higher. We assume two opacity models in the tests: $\lambda = 0.04 \, \mathrm{mag \, Gpc}^{-1}$ in model A and $\lambda = 0.02 \, \mathrm{mag \, Gpc}^{-1}$ in model B. These values are lower than for the SN Ia data, because the SN Ia luminosity is measured in the B band, while the GRB or quasar luminosities are based on bolometric observations. In model A, the errors in $D_L$ and $D_A$ simulate uncertainties in currently available datasets: $\pm1.1$ mag in the $D_L$ distance modulus deriving from GRBs (Fig.~\ref{fig:2}), and $\pm60\%$ in $D_A$ deriving from the ultracompact RSs (Fig.~\ref{fig:4}). In model B, we consider twice-lower uncertainties in $D_L$ and $D_A$ to anticipate improved observations in the future. The datasets are formed by 500 points, which cover uniformly the whole redshift interval $0 < z < 5$.

Fig.~\ref{fig:8} shows the simulated $D_L$ and $D_A$ data in models A and B together with optical depth $\tau(z)$ calculated using Eq. (3) in redshift bins. The figure indicates that the opacity of $0.04 \, \mathrm{mag \, Gpc}^{-1}$ is detectable for sufficiently dense data with uncertainties of current measurements. If the accuracy of observations increases twice, even the opacity of $0.02 \, \mathrm{mag \, Gpc}^{-1}$ might be recognized, particularly at redshifts $z > 3$. Fig.~\ref{fig:8} also demonstrates that differences between the opaque EdS model and the $\Lambda$CDM model are tiny in $D_L$ but well distinguishable in $D_A$. For $z \gtrapprox 0.6$, both $D_A$ curves are similar but with a visible and roughly constant offset. This points to the essential importance of valid $D_A$ data with $z \lessapprox 0.6$ and to the danger of reducing the analysis to data with $z \gtrapprox 0.6$. If  $D_A$ data are used at $z \gtrapprox 0.6$ only \citep{Cao2017, Cao2018, Liao2019}, the $D_A$ offset must be determined by some calibration and any information provided by the $D_A$ data is lost in this way. 

We performed also the other tests analogous to those in the analysis of the low-redshift data. We assumed $D_A$ data biased due to an implicit assumption of the $\Lambda$CDM model in the calibration process and calculated the optical depth $\tau(z)$. As in the low-redshift case, the tests incorrectly indicated no opacity for both datasets corresponding to models A and B.

\subsection{Gravitational waves}

\begin{figure*}
\includegraphics[angle=0,width=16cm,trim=60 200 40 80, clip=true]{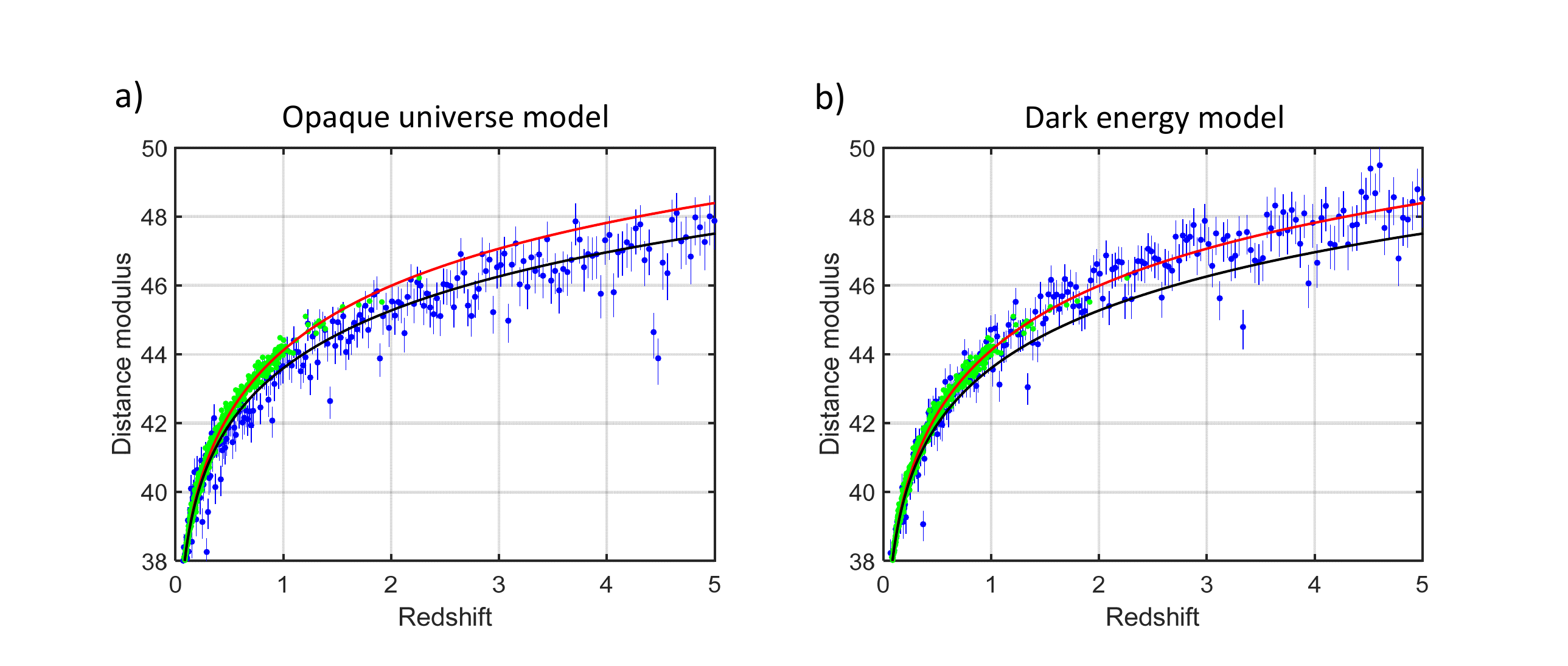}
\caption{
Synthetic high-redshift $D_L$ data simulating anticipated 500 GW observations with expected uncertainties defined by signal-to-noise ratio $\rho = 10$ and weak lensing  $\sigma^{\mathrm{lens}}/D_L = 0.05 z$. The GW data are calculated for: (a) the opaque EdS model (black line), and (b) the $\Lambda$CDM model (red line). Green dots - the SN Ia Pantheon data.  
}
\label{fig:9}
\end{figure*}

In future tests, an alternative measure of the opacity-free luminosity distance might be given by gravitational waves (GW) as the standard sirens \citep{Holz_Hughes2005, Wolf_Lagos2020}. In this case, Eq. (3) can be modified as follows
\begin{equation}\label{eq8}
\frac{D_L^{obs}}{D_L^{GW}} = e^{\tau(z)/2} \, ,									
\end{equation}
where $D_L^{obs}$ is the observed luminosity distance potentially affected by cosmic opacity and $D_L^{GW}$ is the luminosity distance determined from the GW data. These data will be also a powerful tool for studying the universe expansion and cosmology \citep{Taylor_Gair2012}.

The GW data provided by the Einstein Telescope (ET), the planned third-generation GW detector, will be 10 times more sensitive than current ground-based detectors and could detect neutron star-neutron star mergers up to $z \approx 2$ or black hole-neutron star mergers up to $z \approx 5$. According to \citet{Cai_Yang2017}, 102 GW sources with accurately determined redshifts might be detected per year. Potential errors in measurements comprise lensing uncertainties $\sigma^{\mathrm{lens}}$, estimated as $\sigma^{\mathrm{lens}}/D_L = 0.05 z$, and instrumental uncertainties $\sigma^{\mathrm{inst}}$, which depend on the signal-to-noise ratio $\rho$ as  $\sigma^{\mathrm{inst}}/D_L = 2/\rho$. Obviously, a larger chirp mass increases $\rho$ and leads to smaller errors in the distance measure \citep{Cai_Yang2017, Qi2019}.

Several authors simulated the GW data and compared their accuracy with the SN Ia data or GRBs and their potential in opacity tests via the DDR. For example, \citet[their fig. 1]{Yang2019} generated three mock GW datasets with 600, 900 and 1200 points for redshifts up to 5. The simulations indicate that the errors in the GW data are higher than those for the SN Ia data, but smaller than for the GRBs or quasars \citep{Liao2019, Yang2019, Zhou2019}. As shown in Fig.~\ref{fig:9}, the expected accuracy of the GW data detected by the ET should provide unambiguous and opacity-free evidence of the validity of the $\Lambda$CDM model and quantify the opacity effects in the SN Ia data.

\section{Discussion}

The DDR is viewed as a powerful tool for testing for the opacity of the universe, which should work independently of any cosmological model. It was applied by many authors and became quite popular. Nevertheless, a thorough analysis reveals that applying the DDR in cosmic opacity tests is tricky. The applicability of the DDR is strongly limited because of a rather low accuracy and deficient extent of currently available data. At present, the highest accuracy is achieved for the luminosity distance $D_L^{\rm obs}$ using large SN Ia compilations such as Union2.1 or Pantheon \citep{Sullivan2011, Suzuki2012, Scolnic2018}. However, as shown by \citet{Lima2011} and \citet{Vavrycuk2019}, the interpretation of the SN Ia data is not unique and the SN Ia luminosity can be fitted equally well by the transparent $\Lambda$CDM as well as opaque EdS models (for the Pantheon dataset, see Fig.~\ref{fig:1}a,b). Hence, dark energy in the $\Lambda$CDM model produces the same effect as cosmic opacity in the EdS model.

Hence, the resolution of the opacity tests depends primarily on the $D_A$ data, which are however full of pitfalls. First, $D_A$ is measured with a considerably lower accuracy \citep{Jee2015} than $D_L^{\rm obs}$ obtained from the SN Ia observations. Second, some methods like BAO need the statistics of a large number of galaxies and the number of data points is very limited. Third, many approaches are not completely model independent and assume the standard $\Lambda$CDM cosmology in the $D_A$ calculations. In such cases, the zero opacity produced by the DDR might be false being an artefact of the circularity problem (Figs~\ref{fig:6}d,~\ref{fig:7}d). Fourth, $D_A$ is strongly redshift dependent for $z < 0.8$. For higher $z$, its redshift dependence is weak and lost in data scatter. Hence, accurate $D_A$ data for $z < 0.8$ are essential in opacity tests, but such data are mostly missing or inaccurate (e.g., ultracompact RSs).

Cosmic opacity tests via the DDR frequently suffer from other drawbacks and flaws:

\begin{itemize}
\item{
Most authors parametrize the cosmic opacity by a prescribed phenomenological function of redshift, see Eqs (4) and (5). Hence, the validity of such tests is merely limited  to showing consistency or inconsistency that cosmic opacity follows this function. No general conclusion about the transparency of the universe can be deduced from such tests. This is documented by contradictory results obtained for different opacity parametrizations \citep{Lima2011, Holanda2017, Li_Lin2018, Ma2019}. 
}
\item{
Some authors use GRBs calibrated by SN Ia data \citep{Kodama2008, Demianski2017, Holanda2018}. However, such an approach is incorrect, because cosmic opacity, if present, depends on wavelength according to the extinction law \citep{Mathis1990, Li_Draine2001, Draine2003} and it attains different values for different types of data. Hence, if the opacity is non-zero, the GRBs or quasars cannot be calibrated by SNe Ia, because they are affected by opacity in a different way.
}

\item{
In order to tighten constraints on cosmic opacity or parameters of cosmological models, some authors fuse various $D_L$ datasets. For example, SN Ia data are mixed with GRBs \citep{Fu_Li2017}, with quasar data \citep{Risaliti_Lusso2019} or with GRBs and quasar data \citep{Lusso2019}. Even though the $D_L$ datasets are correctly calibrated, their mixing in the DDR must be avoided, because it is physically wrong. In fact, it means that we try to find a frequency-independent cosmic opacity. Obviously, no such opacity can be found and the DDR test must fail and apparently yield a zero opacity. 
}

\item{
Since the $D_L$ datasets might be differently sensitive to the opacity, we have to be careful about generalizing results about the opacity obtained for specific data.  For example, using GRBs in testing for cosmic opacity produced by cosmic dust is rather controversial, because the GRBs are so highly energetic events that their photons can destroy dust grains instead of being absorbed by them \citep{Draine_Hao2002, Morgan2014}. Hence, the GRB observations reflect physically different processes than simple luminosity dimming due to dust absorption of low-energy photons.
}
\end{itemize}

As a consequence, no convincing evidence concerning the transparency of the universe using the DDR has so far been reported. For future studies, it is more convenient to avoid combining the $D_L$ and $D_A$ data via the DDR and to focus rather on their separate analysis. The $D_L$ data can be used for preselecting acceptable cosmological models of a transparent and opaque universe. Then, the optimum cosmological model can be found by fitting with the $D_A$ data or other opacity-free data such as gravitational waves. The cosmic opacity tests should be applied to individual redshift bins independently \citep{Ma_Corasaniti2018}, without confining the opacity to some a priori specified redshift dependence. Also, the $D_L$ and $D_A$ data should be carefully checked to be independent of the $\Lambda$CDM model, otherwise the main claimed strength of the DDR as a powerful cosmology-independent tool is lost. Finally, any result of the DDR test for cosmic opacity will have no general validity, but it will characterize just the specific frequency range of the $D_L$ data used in the test.

%
\section*{Data availability} 
No new experimental data were generated or analysed in support of this research. The synthetic data underlying this article will be shared on reasonable request to the corresponding author.


\bibliographystyle{mnras}

\bibliography{paper} 

\end{document}